\documentclass[%
 reprint,
 amsmath,amssymb,
 aps,
 groupedaddress,
 prl
]{revtex4-2}
\usepackage{graphicx}
\usepackage{dcolumn}
\usepackage{bm}
\usepackage[version=3]{mhchem}
\usepackage{csquotes}
\usepackage{hyperref}
\hypersetup{
    colorlinks=true,
    linkcolor=blue,
    filecolor=magenta,
    urlcolor=cyan,
}
\usepackage{color}

\begin{document}
\preprint{PRL}
\title{Vibrational density of states capture the role of dynamic allostery in protein evolution}

\author{Tushar Modi}
\affiliation{Department of Physics, Arizona State University, Tempe, Arizona 85287, USA}

\author{Matthias Heyden}
\affiliation{School of Molecular Sciences, Arizona State University, Tempe, Arizona 85287, USA}
\affiliation{co-corresponding}

\author{S. Banu Ozkan}
\affiliation{Department of Physics, Arizona State University, Tempe, Arizona 85287, USA}
\affiliation{co-corresponding}

\date{\today}

\begin{abstract}
Previous studies of the flexibilities  of ancestral proteins suggests that proteins evolve their function by altering their native state ensemble. Here we propose a more direct method of visualizing this by measuring the changes in the vibrational density of states (VDOS) of proteins as they evolve. Through analysis of VDOS profiles of ancestral and extant proteins we observe that $\beta$-lactamase and thioredoxins evolve by altering their density of states in the terahertz region. Particularly, the shift in VDOS profiles between ancestral and extant proteins suggests that nature utilize dynamic allostery for functional evolution. Moreover, we also show that VDOS profile of individual position can be used to describe the flexibility changes, particularly those without any amino acid substitution.
\end{abstract}

\maketitle

\section{Introduction}
Allostery  uses long distance interactions between different regions of a protein to  orchestrate the unique dynamics necessary for its function. Classically, it was described as a phenomenon whereby a ligand binding event at a distal site affects the binding activity of a functional site through conformational changes, allowing remote functional regulation~\cite{monod_nature_1965, changeux_allostery_2012}. It has been shown to be of prime importance in regulating several biological processes such as metabolism~\cite{link_advancing_2014}, transcription~\cite{wright_intrinsically_2015, li_genetically_2017}, signal transduction~\cite{changeux_allosteric_2005, changeux_allostery_2012, falke_architecture_2014, nussinov_underappreciated_2013}, etc. However, the general mechanism of allostery goes beyond just ligand binding. Studies suggest that the modulation of local interactions, not merely by ligand binding, but also by other phenomenon like mutations at residue positions distal from the active site could affect binding or enzymatic activity~\cite{mcleish_allostery_2013, wodak_allostery_2019, doi:10.1063/5.0011392, doi:10.1063/5.0039175, 10.1371/journal.pone.0199225}.

The current understanding of allostery is derived from the ensemble model of proteins~\cite{wodak_allostery_2019} inspired from the Cooper's model of dynamic allostery~\cite{cooper_allostery_1984}. It describes how allosteric mechanism does not need to rely only on distinct conformational change, but modulation of normal modes leading to shifts in native ensemble are enough to alter the underlying energy landscape, thereby, regulating the function. Recent studies on reconstructed ancestral proteins also suggest that nature may use the same dynamic allostery principles for protein evolution. Particularly, comparative computational studies of ancestral proteins and their modern homologs have shown that during evolution, mutations, mostly at regions far from the functional site, alter the conformational dynamics of the proteins for the emergence of a new function or fine-tuning their pre-existing functions while conserving their 3D fold~\cite{modi_hinge-shift_2021, modi_tushar_ancient_2018, kim_hinge_2015, campitelli_role_2020, zou_evolution_2015, modi_hinge-shift_2021}.

In our previous studies, we have used our protein-dynamics based metric, Dynamic Flexibility Index (DFI)~\cite{nevin_gerek_structural_2013} to analyze how changes in the conformational dynamics modulate the function during evolution of several different type of protein systems such as Green Fluorescent proteins (GFP)~\cite{kim_hinge_2015}, $\beta$-lactamase~\cite{modi_protein_2021, zou_evolution_2015} and Thioredoxin (Thrx)~\cite{modi_tushar_ancient_2018}. DFI is a position specific metric which uses linear response theory to analyse MD trajectories to  quantify  the relative flexibility of a residue position with respect to the rest of the protein. Using DFI analysis, we have shown that proteins adapt to new environments or enhance their enzymatic activity through hinge shift mechanism such that the flexibility profile associated with their function is altered through point mutations. Particularly, rigidification of some sites are compensated by enhancements in flexibility of other distal rigid sites ~\cite{campitelli_role_2020,modi_hinge-shift_2021,modi_protein_2021}. Furthermore, during evolution, rather than directly mutating the active site residues, distal mutations modulate the flexibility of the functional sites to accommodate novel non-cognate substrate degradation while conserving the 3-D fold~\cite{campitelli_role_2020, modi_hinge-shift_2021, modi_tushar_ancient_2018, ijms19123808}. The mutations utilize the anisotropic nature of the 3D network of interactions between different residues in proteins to alter the relative population of the accessible conformations in their native state ensemble~\cite{campitelli_role_2020, modi_hinge-shift_2021, modi_tushar_ancient_2018}. Indeed, this is nothing but similar to the principle of dynamic allostery where the allosteric regulation by ligand binding (i.e., changes in network of interactions upon binding) is shown to be achieved through modulation of the native state ensemble without having any observable effect on the native state conformation.

In this study, we aimed to bridge the gap between observed changes in the DFI profile of a protein and a corresponding shift in the native state ensemble through the lens of vibrations in a protein (using vibrational density of states~\cite{near_deng,Turton2014TerahertzUV}). This would provide a first-hand proof of the utilization of the dynamic allostery by nature for evolution. In addition, we also attempted to analyse how vibrational spectra change for the residue positions which have preserved their dynamics as well as have underwent dynamical changes during evolution. Therefore, we showed how the VDOS analysis provides a microscopic vision at the residue level and at different time scales previously not accessible through DFI.

\section{Results and Discussion}
As discussed in the previous section, mutations inducing changes in the DFI profile of a protein describe how these fine-tune the function through alteration of dynamics while conserving the 3-D fold. DFI analysis suggests that these mutations mould the native state ensemble for the evolution of proteins.  In order to investigate this further,  we calculated the vibrational density of states (VDOS) of the ancestrally reconstructed and modern proteins belonging to two families of enzymes, $\beta$-lactamase and Thrx enzymes. Extensive studies have been performed by us in the past on these enzymes, where, using the DFI analysis, we have described the critical role of protein dynamics in their evolution~\cite{modi_hinge-shift_2021, modi_tushar_ancient_2018,ijms19123808}. 

Here, in order to compute the VDOS for these enzymes, we ran 1 ns micro-canonical (NVE) molecular dynamics simulations for both, the ancestral and extant enzymes (for $\beta$-lactamases and Thrx). The starting points for these simulations were selected based on the conformations highly visited during longer simulations of each enzyme under canonical (NPT) ensemble. Afterwards, using the NVE simulation trajectory, we calculated velocity auto-correlation functions of length 4 ps for the $C_{\alpha}$ atom of each residue position in the enzyme. Then the Fourier transform of the velocity auto-correlation functions was used to calculate the VDOS of the $C_{\alpha}$ atom of each residue position in the enzyme (for details on the time scales and calculations see Supplementary information). The VDOS for the residue positions provides the information of the vibrational modes accessible to the amino acid.

In addition, to be able to qualitatively compare the native state ensemble of the ancestral and extant enzyme through differences in their vibrational dynamics, we also computed the effective VDOS of the enzyme using the mean velocity auto-correlation function averaged over the velocity auto-correlation functions of the $C_{\alpha}$ atoms of each residue position. For further details on the calculation of VDOS, please refer to the supplementary information.

\subsection{Vibrational spectra sheds light on the modulation of native state ensemble during protein evolution.}
\begin{figure}[h!]
\centering
\includegraphics[width=8.6cm]{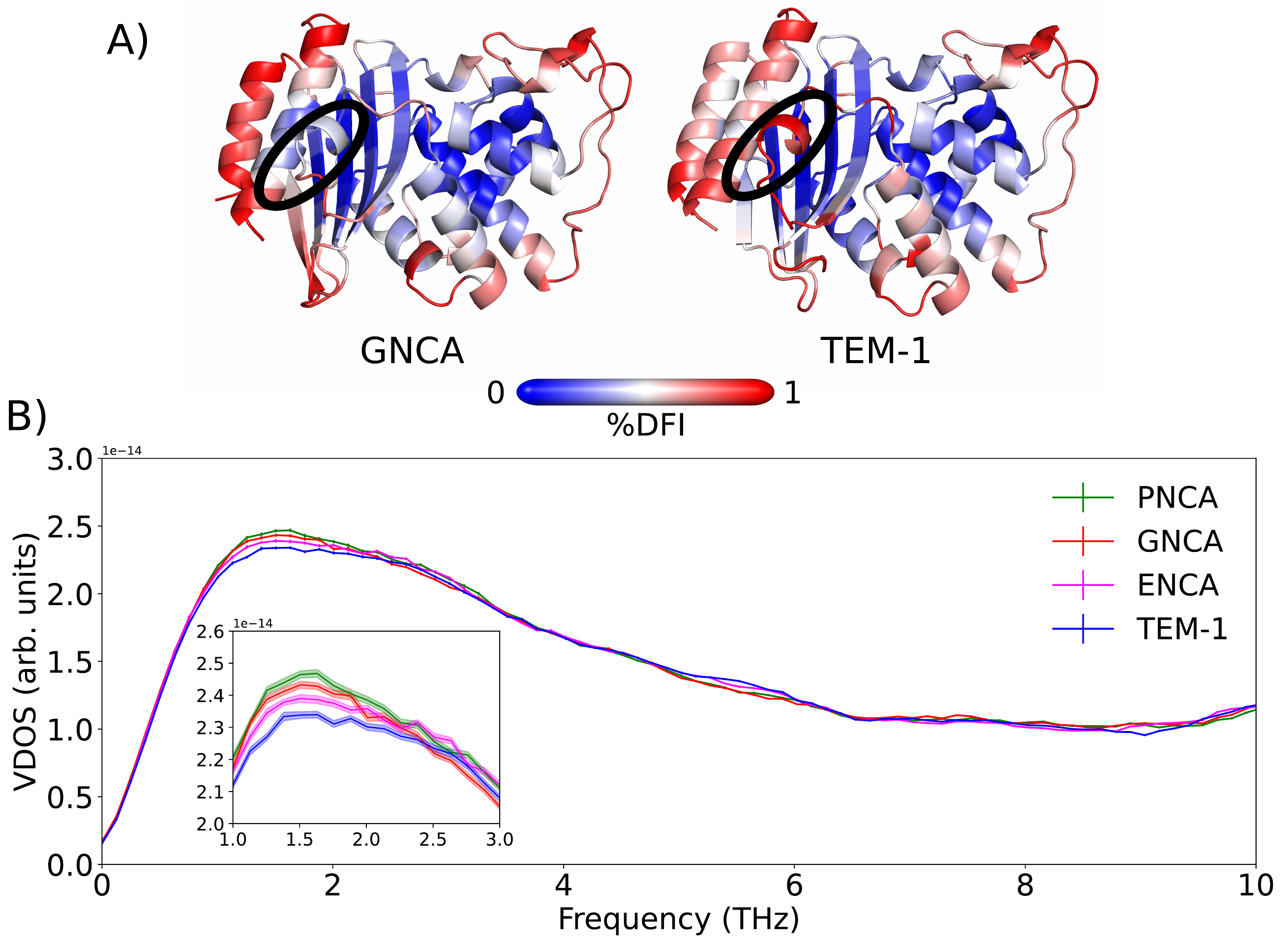}
\caption{\label{fig:1} \textbf{Evolution alters the native state ensemble of the enzymes.} Mutations accumulated through evolution fine-tune the function of the enzyme by altering the native state ensemble and hence the flexibility as shown by DFI profile in (A) for $\beta$-lactamase enzymes (ancestral GNCA, and extant TEM-1). These describe how the flexibility profile of the proteins evolved as shown by the comparison of their cartoon representation color coded according to their percentile ranked DFI profiles. This change can also be observed in (B) by comparing the vibrational density of states (VDOS) of the ancestral $\beta$-lactamases (PNCA, GNCA and ENCA in the order of their divergence time) with the extant $\beta$-lactamase (TEM-1). The differences in the VDOS in lower frequencies are also highlighted in the inset panel where the shaded region show the standard error of mean.}
\end{figure}

The VDOS of the ancestral and extant enzymes of $\beta$-lactamase (Figure~\ref{fig:1}) and Thrx family (Supplementary Information Figure S1 and S2) exhibit significant differences indicating the change in their native state ensemble during evolution in the form of changes in population between different conformations in the ensemble. Particularly, we observed distinct peak shifts in their VDOS around the lower (1-2 THz) frequency regions Figure~\ref{fig:1}B. These lower frequencies describe motion at slower time scales critical for function (also highlighted in Figure~\ref{fig:1}).

The ancestral Precambrian $\beta$-lactamases (the Gram-negative bacteria ancestor, GNCA and Gram-positive and Gram-negative bacteria ancestor, PNCA) differ from their modern counterpart in their function. The ancestors are moderately efficient promiscuous enzymes capable of degrading a diversity of $\beta$-lactam antibiotics in contrast with the  modern TEM-1 and its closer ancestral homolog of enterobacteria, ENCA  $\beta$-lactamase, which are specialist degrading only penicillin with higher efficiency. This generalist to specialist enzyme evolution involved more than 100 mutational changes which alter the flexibility profile of the enzyme (Figure~\ref{fig:1}A), while, interestingly, conserving the 3-D fold and the catalytic residues. In Figure~\ref{fig:1}B, we observed that PNCA and GNCA $\beta$-lactamase have a higher VDOS at 1.5 THz as compared to TEM-1 and ENCA $\beta$-lactamases. This evidence supports our previous findings~\cite{ijms19123808,modi_hinge-shift_2021,modi_tushar_ancient_2018} that the mutations at distal sites uses dynamic allostery to shape the native state ensemble to modulate substrate specificity. Similar results are observed in other studies where the native state ensemble, which governs the vibrational motions in a protein, have been shown to also regulate its binding interactions with ligands~\cite{Turton2014TerahertzUV,PhysRevLett.93.028103,Niessen2017MovingIT}.

Similar shifts in the VDOS were observed in the enzymes from Thrx family, which have evolved their functions to adapt at a lower temperature as well as lower acidic conditions (higher pH) while conserving the 3D fold~\cite{modi_tushar_ancient_2018}. A shift in the peak of VDOS around 1 THz was observed when ancestral and extant Thrx enzymes in human and bacterial branches were compared (Supplementary Information, Figure S1 and S2). In AECA Thrx (the human ancestor), the VDOS has a peak at ~1.1 THz which, upon evolution, has shifted to a higher frequency of ~1.5 THz in LECA, LAFCA and Human Thrx. Moreover, the extant Thrx in Human branch has the largest shift of all. In bacterial branch of Thrxs, on other hand, in LBCA Thrx (the bacterial ancestor), the peak in lower frequency at 1.1 THz has also shifted towards a higher frequency of ~1.5 THz in LPBCA and LGPC Thrx. These shifts in low frequency range support the role of dynamic allostery in evolution. Due to the role played by lower frequencies in regulating slower and global motions, the presence of a peaks in VDOS at lower frequencies in ancestral enzymes also suggests that these have a larger conformational diversity. This finding is also supported by our previous studies on ancestral enzymes~\cite{modi_hinge-shift_2021,modi_tushar_ancient_2018,ijms19123808,campitelli_role_2020}. However, the extant E.coli Thrx does not follow this trend and comes up as an outlier with its peak at 1.5 THz but also having a spectra with lower frequencies more populated than its ancestors. This agrees with our other studies where E.coli Thrx has been, once again, labelled as an outlier when compared with other Thrx enzymes in their phylogenetic tree due to its very low catalytic rate and thermal stability~\cite{modi_tushar_ancient_2018}.

\subsection{VDOS captures the shift in the dynamics of protein residue positions through evolution.}
As described earlier, the DFI analysis has shown that proteins adapt to a newer environment or fine-tunes their pre-existing function through the hinge-shift mechanism where the increased flexibility of some rigid (i.e., hinge) sites is compensated by the rigidification of other sites~\cite{campitelli_role_2020,modi_hinge-shift_2021,kim_hinge_2015}. Therefore, we further analyzed the VDOS of the $C_{\alpha}$ atoms of each residue position in the protein chain to explore whether the comparison of VDOS for individual positions could illustrate the dynamical changes. Thus, capturing the functional evolution and providing an additional probe for observing the altered native state ensemble.

First, we focused on the residue positions which have not underwent any mutations during evolution and have also conserved their rigidity (i.e., maintained a lower \%DFI score of less than 0.2 as a hinge in ancestral and extant protein). Such residue positions have a higher significance for orchestrating motions throughout the protein. Comparing their VDOS, we observed that, as expected, such residue positions preserve their spectra in lower as well as higher frequency regime (for $\beta$-lactamases, see Figure~\ref{fig:2} and for Thrx, see Supplementary information Figure S3).

\begin{figure}[h!]
\centering
\includegraphics[width=8.6cm]{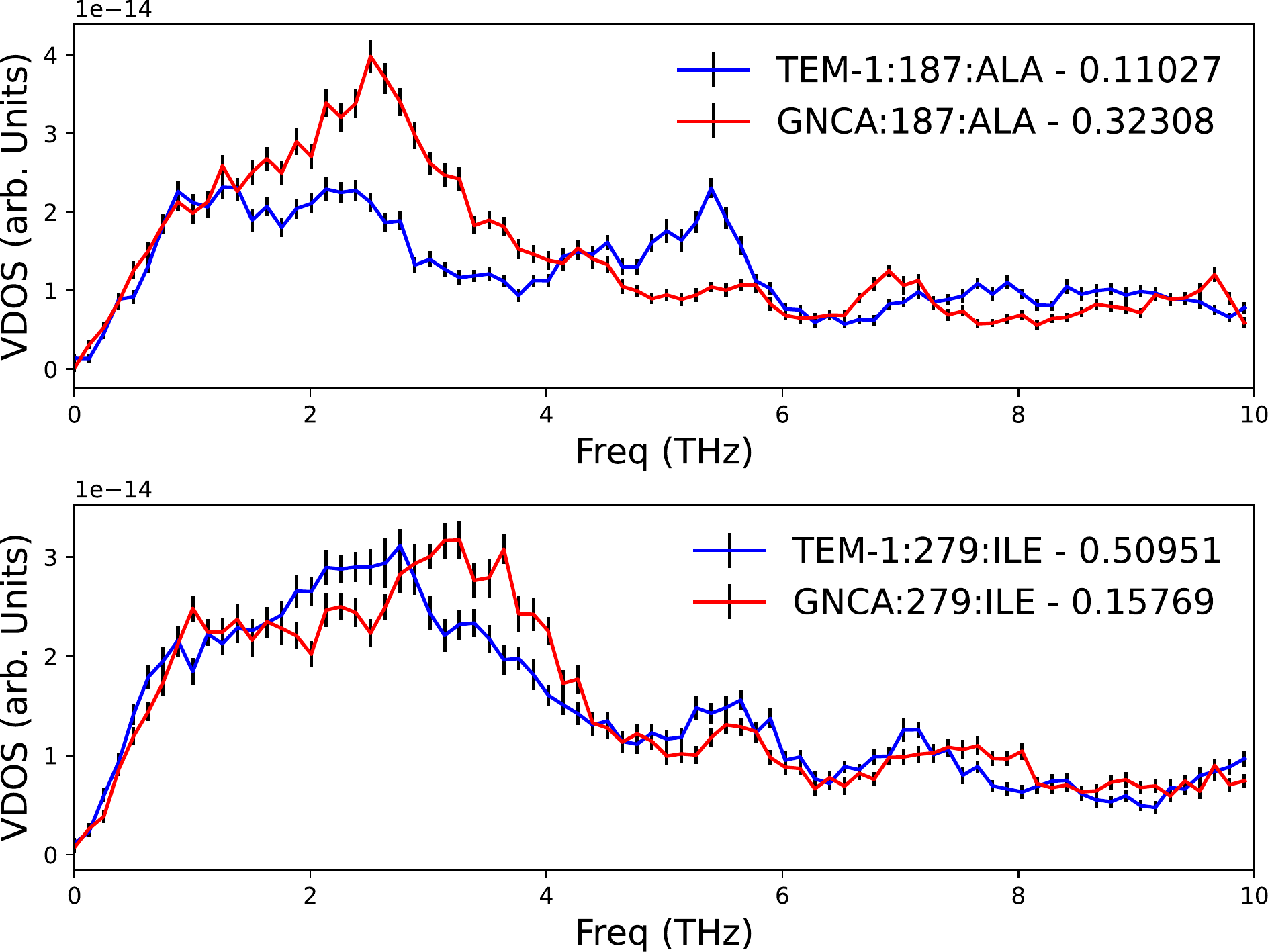}
\caption{\label{fig:3} \textbf{Change in VDOS of the residue positions with no mutation but a change flexibility through evolution in GNCA and TEM-1 $\beta$-lactamases.} Here we observe a subset of the residue positions (187 top and 279 bottom) with a conserved amino acid but a change in their \%DFI score. Upon comparison of their VDOS revealed differences, particularly in lower frequencies where an increase in flexibility was accompanied by a shift of lower frequency peaks towards even lower frequencies. In addition, the inverse of this was also observed. The black bars in the plots represent the standard error of the mean.}
\end{figure}

Second, we compared the VDOS of the residue positions in $\beta$-lactamases with a major change in their flexibility profile despite having no mutations between GNCA to TEM-1 $\beta$-lactamase. These are the positions which have either lost their flexibility, thereby becoming a rigid hinge (i.e., a residue position with \%DFI\textsubscript{GNCA} $>$ 0.3 and \%DFI\textsubscript{TEM-1} $<$ 0.2) or have gained flexibility (i.e., a residue position with \%DFI\textsubscript{GNCA} $<$ 0.2 and \%DFI\textsubscript{TEM-1} $>$ 0.3). Through the analysis, we observed that such residue positions which have gained flexibility, typically exhibit changes in their VDOS such that the peaks at low frequencies have further shifted towards even lower frequencies. As a result, making the lower frequencies more populated (Figure~\ref{fig:3}). On the other hand, the inverse is also true where we observed that the residue positions which have lost their flexibilities have their lower frequency peak shifted towards a higher frequency (Figure~\ref{fig:3}). These findings support our hypothesis that the lower frequencies, which dominate the slower and global motions in proteins, make the residue positions more flexible (such that, it can sample a larger amount of conformational space). Such motions are also predominantly captured by the DFI metric which samples at a significantly slower time scales corresponding to tens of nanosecond as opposed to time scales of the order of tens of picoseconds to nanosecond which account of lower frequencies in the VDOS. This behavior was also observed in the residue positions in Thrx from bacterial and human lineages which have undergone flexibility shifts due to evolution despite being conserved (see Supplementary Information Figure S4).

\begin{figure}[h!]
\centering
\includegraphics[width=8.6cm]{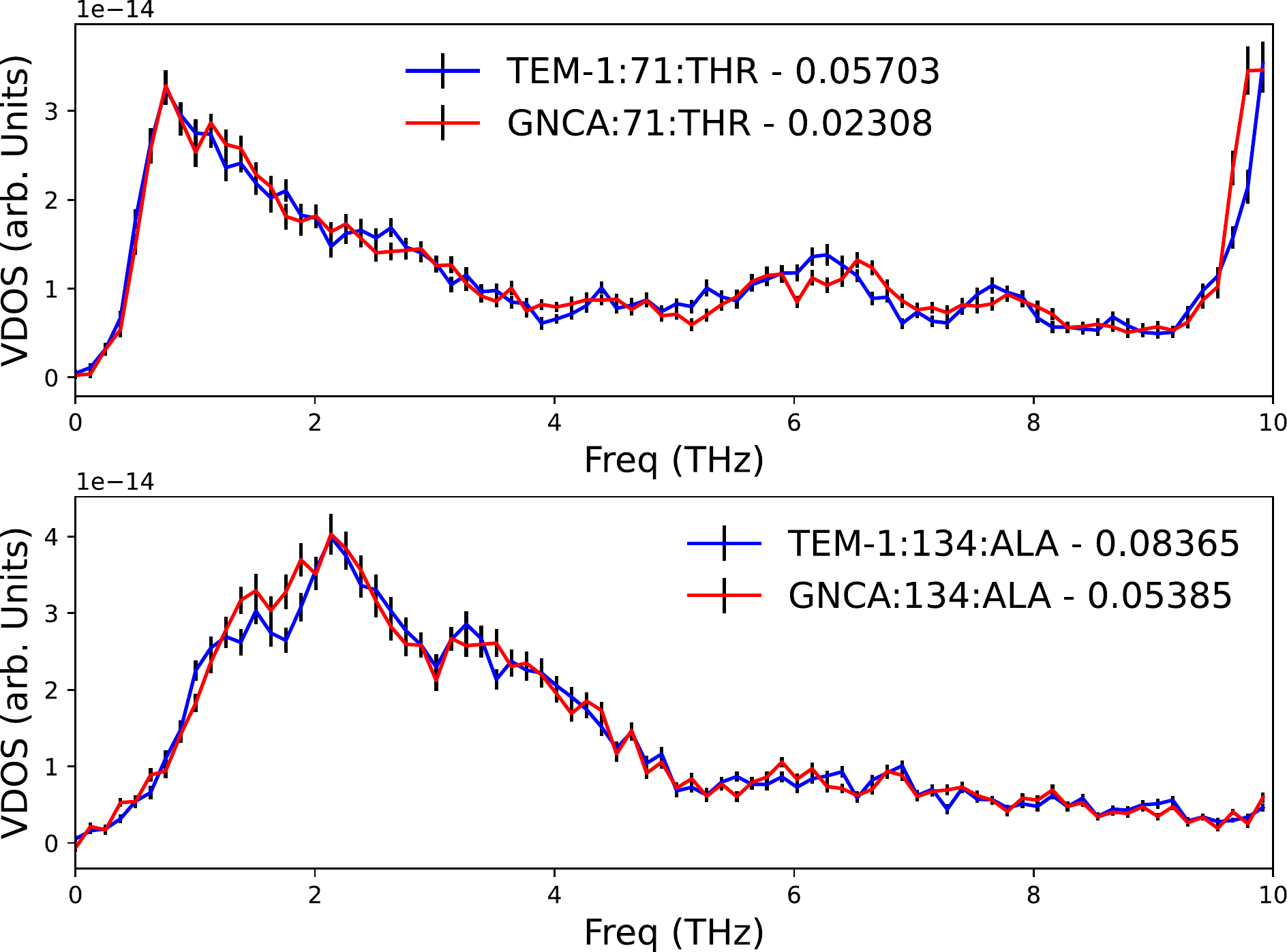}
\caption{\label{fig:2} \textbf{Evolution preserves the VDOS of the residue positions with no mutation and no change in flexibility in $\beta$-lactamases.} Here we observe a subset of the residue positions (71 top and 134 bottom) with a conserved amino acid and a very low \%DFI score. Upon comparison of their VDOS also showed a significant amount of similarity, particularly in lower frequencies ($<$5 THz). This  reveals that these dynamically conserved positions evolved while preserving their VDOS as well. The black bars in the plots represent the standard error of the mean.}
\end{figure}

However, visually comparing the differences in the lower frequency regime of the VDOS for ancestral and extant enzymes, might not provide a complete description of the dynamical changes during evolution (see Supplementary Information Figure S5). Therefore, in order to check whether incorporating the contribution of all the frequencies sampled in the calculation of VDOS would provide a clearer picture of the flexibility of a residue position, we calculated the entropic contribution of each residue position from their respective VDOS. We modeled each residue position as being composed of independent harmonic oscillators, such that their VDOS would describe the distribution of the frequencies at which they would oscillate (i.e., their density of states). In this model, for each residue position (for $C_{\alpha}$ atom), we then compute the partition function which can be used to calculate the entropy (further details of the calculations are provided in the supplementary information).

\begin{figure}[h!]
\centering
\includegraphics[width=8.6cm]{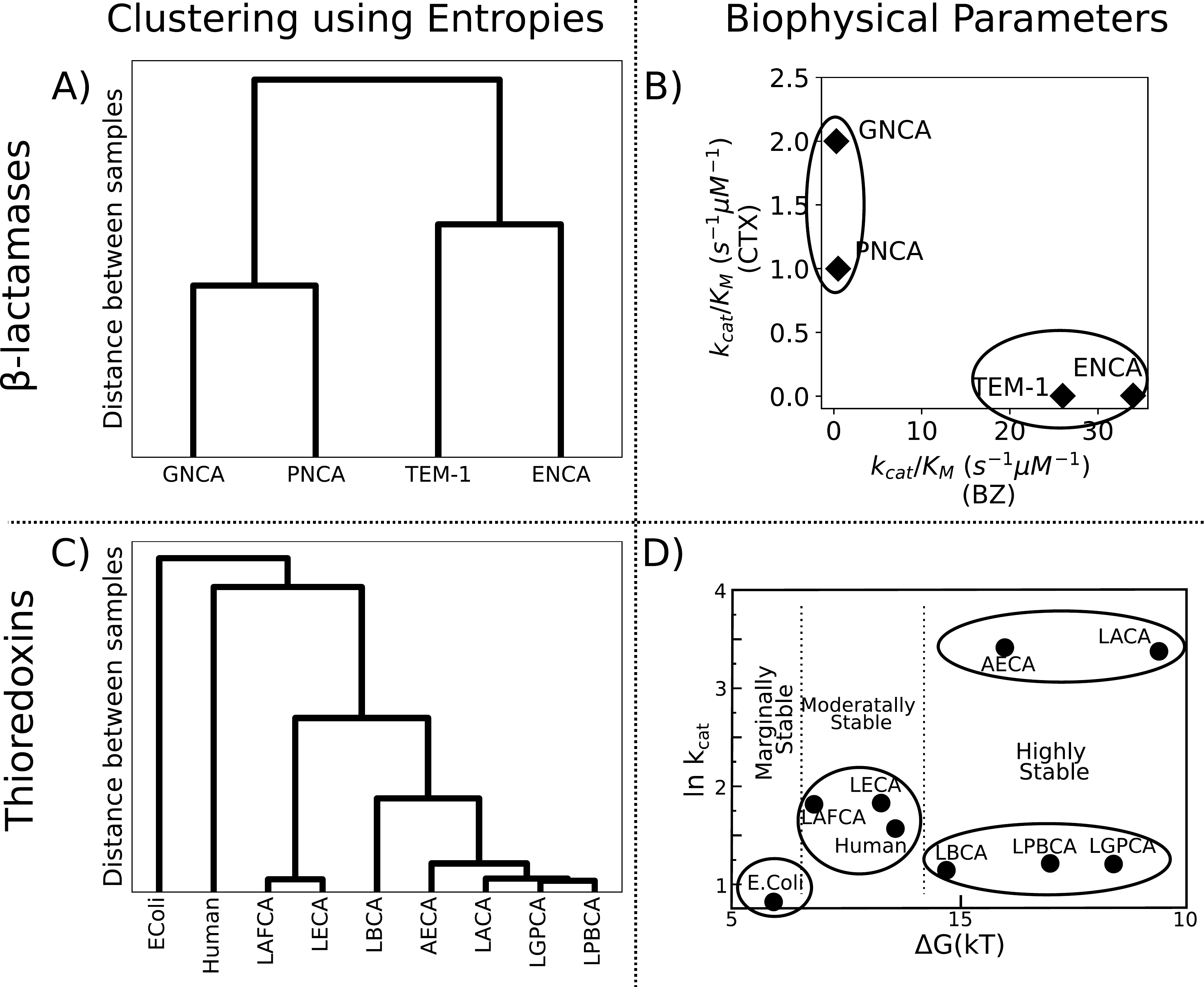}
\caption{\label{fig:4} \textbf{Clustering ancestral and extant proteins from $\beta$-lactamase and Thrx phylogenetic tree according to their entropic profiles.} Clustering the enzymes based on their VDOS revealed that the groups so obtained, segregate the enzymes based on their experimentally obtained biophysical properties.}
\end{figure}

Utilizing this approach, we computed the residue specific entropy profile for the ancestral and extant $\beta$-lactamases (PNCA, GNCA, ENCA and TEM-1) and  also  for the Thrxs containing the ancestral and extant proteins from bacterial (LBCA, LGPCA, LPBCA, E.coli), Archea (AECA, LACA) and Human (LAFCA, LECA, Human) branches of the phylogenetic tree. Using the residue specific entropic profile for each protein as features, we then hierarchically clustered the enzymes together (using their lowest three principal components obtained through SVD). Thereafter, we observed whether the entropic profiles calculated via VDOS were able to group the proteins based on their functionally relevant properties like activity and/or stability (Figure~\ref{fig:4}).

In our previous studies~\cite{modi_hinge-shift_2021, modi_tushar_ancient_2018, zou_evolution_2015}, it has been shown that the proteins clustered together using their \%DFI profiles share similar  biophysical properties  (i.e., a similarity in their activity and/or stability). Clustering the $\beta$-lactamase enzymes based on their vibrational entropy profiles yielded similar results. The residue specific vibrational entropy profile for each enzyme was able to capture functional differences between then their turnover rates to hydrolyze the antibiotics benzylpenicillin (BZ) and cefotaxime (CTX) ~\cite{zou_evolution_2015} (Figure~\ref{fig:4}A), (Figure~\ref{fig:4}B). It grouped TEM-1 and ENCA $\beta$-lactamases together which are both highly specific in their activity to hydrolyse the antibiotic  benzylpenicillin (BZ) as shown by their very high turnover rates ($\approx$ 20 $s^{-1}\mu M^{-1}$) as opposed to significantly lower turnover rates ($\approx$ 0 $s^{-1}\mu M^{-1}$) to hydrolyze cefotaxime (CTX). On the other hand, GNCA and PNCA $\beta$-lactamases were also clustered together which are both promiscuous in their activity towards these two antibiotics as shown by their moderate activity towards CTX ($\approx$ 1 $s^{-1}\mu M^{-1}$) and BZ ($\approx$ 1 $s^{-1}\mu M^{-1}$). This clustering indeed was identical to the clustering obtained based on their DFI profiles~\cite{zou_evolution_2015}.

Similarly, for Thrx proteins, (Figure~\ref{fig:4}C and D), the entropic profile of each protein clustered them based on the characterized biophysical properties of thermal stability and catalytic rates as captured by DFI profile ~\cite{modi_tushar_ancient_2018}).  Particularly, Thrx from archea (AECA and LACA) and bacterial branch (LBCA, LPBCA and LGPCA) (except E.coli Thrx)  exhibiting high thermal stability were clustered closer to each other. Moreover, AECA and LACA Thrxs which were also sub-grouped together are highly active in their function (based on their disulphide bond reduction rate). On the other hand, the other subgroup LPBCA and LGPCA Thrx have a lower  enzymatic activity.  Human Thrxs (LAFCA, LECA and Human) were all grouped closer to each other according to their entropic profile which aligns with their similar biophysical properties, being moderately active and moderate stability. Finally, E. coli Thrx, was clustered solely as it shares large differences with other Thrx in its entropic profile, is also an outlier according to its biophysical properties, being the least stable and least active Thrx of all (as also shown by our other study on Thrx proteins~\cite{modi_tushar_ancient_2018}). 

The near perfect segregation of ancestral and extant proteins using the entropy calculated from VDOS clearly describes the important role played by motions with intermediate to faster time scales as was also emphasized by other studies performed by us and others~\cite{bahar_vibrational_1998, bahar_structure-encoded_2015, zhang_shared_2019, gobeil_structural_2019}.

\section{Conclusion}
Through VDOS analysis, we provided a first-hand support of the modulation of the native state ensemble through mutations during evolution. This ensemble shift via mutations enables a protein to fine-tune its existing function in order to adapt to changing environmental conditions or acquire new functions as shown by the analysis on ancestral and extant $\beta$-lactamase and Thrx enzymes. In addition, the loss or gain of rigidity by individual residue positions during evolution can be attributed to the peak shift  at lower frequency of VDOS. Therefore, these peak shifts not only provided a more complete mechanistic picture of the changes induced by evolution, but also showed that nature uses dynamic allostery to modulate function.

We also computed the vibrational entropy per position using VDOS of each residue position.  In these calculations, we incorporated the contribution of all the frequencies observed in the VDOS. Through this, we observed that changes in vibrational entropy profiles capture changes in  biophysical characterization. Particularly, the entropic contribution of each residue position modifies with evolution such that the difference between them can cluster the enzymes based on their function and/or thermal stability in the case of $\beta$-lactamase and Thrx enzymes. 

\bibliography{manuscript.bib}
\end{document}


\maketitle

\section{Calculation of Vibrational Density of States (VDOS) from equilibrium MD simulation trajectories}
In order to compute the VDOS for a protein, we used the dynamics sampled via constant energy molecular dynamics simulation of the protein (under conditions of micro-canonical NVE ensemble). The VDOS was computed for each residue separately using the normalized velocity auto-correlation function of the $C_{\alpha}$ atom of the amino acid position as shown below:
\begin{equation}
    g(\omega)=\frac{1}{3k_{B}T}\int_{-\infty}^{\infty}C_{i}(t_{0},t)e^{i\omega t}dt, \label{eq:vdos}
\end{equation}
where, $C_{i}(t_{0},t)$ is the normalized velocity auto-correlation function of the $C_{\alpha}$ atom of the $i^{th}$ residue in the protein chain. 

In order to obtain the velocity auto-correlations of the residue positions, we started with simulating each protein under constant temperature (300K) equilibrium conditions for $1.5 \mu s$ using AMBER molecular dynamics simulation package~\cite{amber} following the protocol described in~\cite{modi_hinge-shift_2021}. We clustered the trajectories using the last $500 ns$ by K-means clustering (dividing the ensemble of conformations into 10 clusters and using a cutoff backbone RMSD of 0.8 \AA). We then used the top two clusters for VDOS analysis. It should be noted that the proteins were simulated for a very long time to ensure that the first two clusters account for more than $98\%$ of the conformations sampled in the last $500 ns$ of the trajectory and the rest of the clusters have an occupancy of less than $1\%$. For each of the two most sampled  clusters, the velocity information is obtained from the equilibrium trajectory (so that their distribution satisfy the Maxwell-Boltzmann distribution for the protein corresponding to $300 K$).

For each cluster representative, we then ran a $1 ns$ constant energy simulation (NVE ensemble) to ensure that the velocity of the atoms are not modified (or re-scaled) by the thermostat algorithms. In this simulation, the velocity information for each atom was stored every $8 fs$. Each trajectory is then divided into 25 equally sized windows and for each window velocity auto-correlation functions of lengths $4 ps$ are calculated as,
\begin{equation}
    C_{i}(t)=\langle \Vec{v_{i}}(0).\vec{v_{i}}(t) \rangle. \label{eq:autocorr}
\end{equation}

For the velocity auto-correlation functions obtained from each of the time windows, Eq.~\ref{eq:vdos} was used to calculate the VDOS. Finally, the VDOS obtained was normalized. In accordance with the Nyquist-Shannon Sampling theorem~\cite{1697831}, the frequency resolution of the spectra (depending on the slowest motion sampled) is $0.125 THz$ and the highest frequency sampled (depending on the save frequency of velocities) is $62.5 THz$. Finally, the VDOS for each residue position was averaged over all the time windows and clusters to give the mean VDOS and the standard error of the mean provided a measure of the fluctuation in the data about the mean.

The effective VDOS of the protein as a whole was calculated using using the mean velocity auto-correlation function averaged over all the $C_{\alpha}$ atoms in the proteins.

\section{Calculation of per residue entropic contribution using their VDOS.}
For the calculation of vibrational entropy of the amino acids at each residue position, we modeled each residue as being composed of a harmonic oscillator which can oscillate at frequencies whose probability distribution is described by the VDOS. For such an oscillator vibrating with a frequency of $\omega$, the partition function can be written as:
\begin{equation}
    Z_{\omega}=\frac{e^{-\beta\hbar\omega/2}}{1-e^{-\beta\hbar\omega}}, \label{eq:Zw}
\end{equation}
where, $\beta = \frac{1}{kT}$. For such a system with $N$ degree of freedom and the VDOS as $g(\omega)$, the net partition function ($Z$) can be given as:
\begin{equation}
    Z_{NET}=\prod_{i=1}^{M}(Z_{\omega_{i}})^{g(\omega_{i})N\Delta\omega},
\end{equation}
where, i is the index of frequencies sampled discretized by discrete Fourier transform and $M$ is the index corresponding to highest frequency sampled. $\Delta \omega$ is the step size in frequency domain. It should be added that during our calculating, we ignore the zero frequency relaxation dynamics of the protein which are not harmonic in nature. This can be used to calculate the Helmholtz free energy (F) of the system as:
\begin{eqnarray}
    F&=&-\frac{1}{\beta}\ln{Z_{NET}} \\
    &=&-\frac{N}{\beta}\sum_{i=1}^{M}\Delta \omega g(\omega_{i})\ln{Z_{\omega_{i}}}. \label{eq:F}
\end{eqnarray}
Using this, entropy for the residue position (S) can be calculated as:
\begin{equation}
    S=-\frac{\partial F}{\partial T} \label{eq:S}
\end{equation}
Using Eqs.~\ref{eq:F} and~\ref{eq:Zw}, Eq.~\ref{eq:S} can be expressed as:
\begin{equation}
    S=kN\Delta \omega\sum_{i=1}^{M}g(\omega_{i})\bigg[\ln{Z_{\omega_{i}}} + \frac{\beta\hbar\omega_{i}}{2}\frac{1}{\tanh{(\beta\hbar\omega_{i}/2})}\bigg].
\end{equation}

\begin{figure}[h!]
    \centering
    \includegraphics[width=8.6cm]{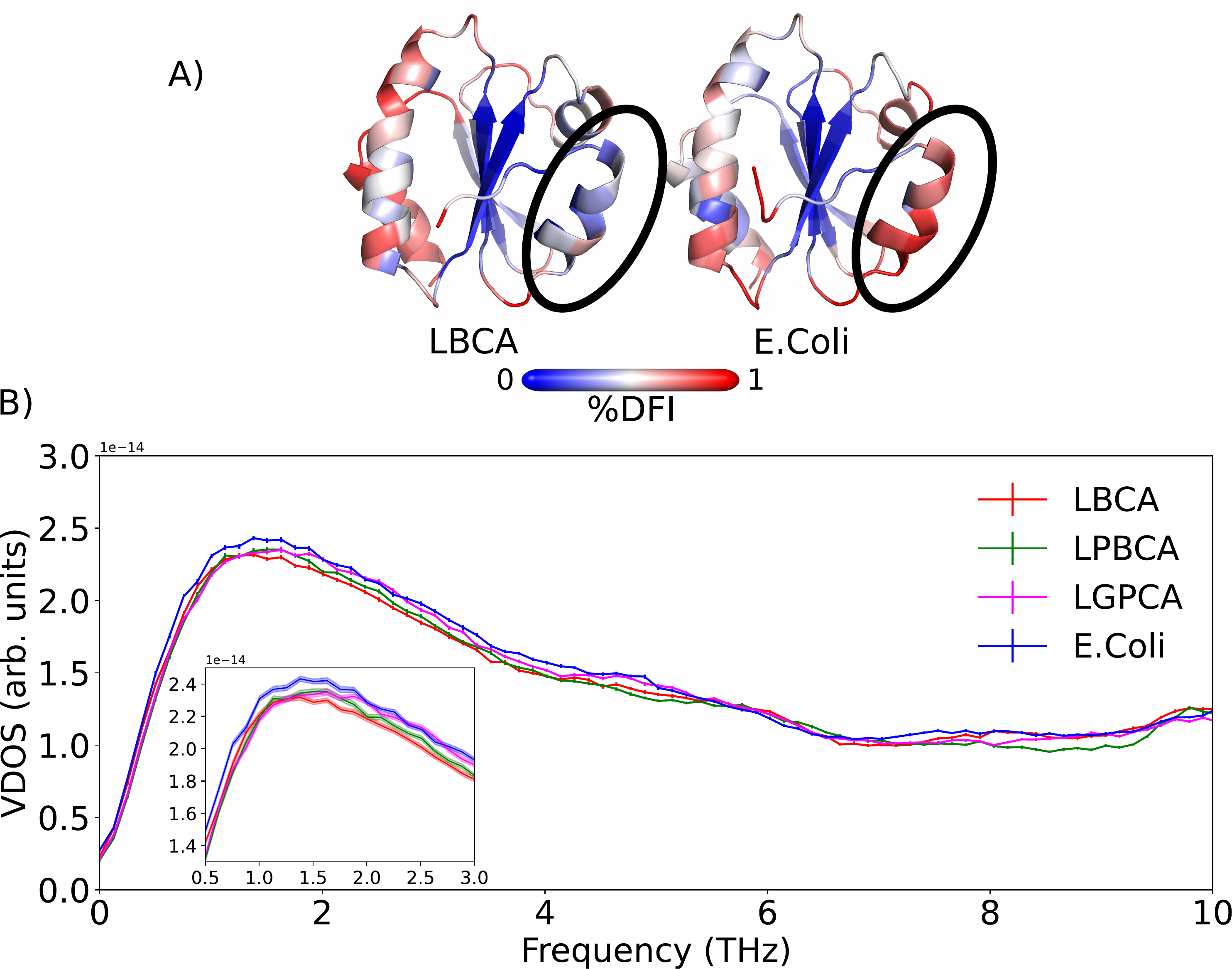}
    \renewcommand{\figurename}{Figure}
    \caption{\textbf{Modulation of the native state ensemble of Thrx enzymes from bacterial branch observed using their VDOS.} Amino acid substitutions accumulated through evolution fine-tune the function of the enzyme by altering the native state ensemble and hence modify the flexibility as shown by DFI profile in (A) for bacterial Thrx enzymes. (A) shows  the flexibility profiles of Thrx enzymes  of the last bacteria common ancestor and E.coli (extant) in cartoon representations color coded according to their percentile ranked DFI profiles within a spectrum of red-white blue  where red shows the highest and blue lowest flexibility. The DFI profiles of ancestral and extant Thrx enzymes shows significant changes in flexibility as shown in circle. This change can also be observed by comparing the VDOS of the ancestral proteins with the extant ones (in (B)). The differences in lower frequencies are focused in the inset where shaded region on the plots show the standard error of mean.}
\end{figure}

\begin{figure}[h!]
    \centering
    \includegraphics[width=8.6cm]{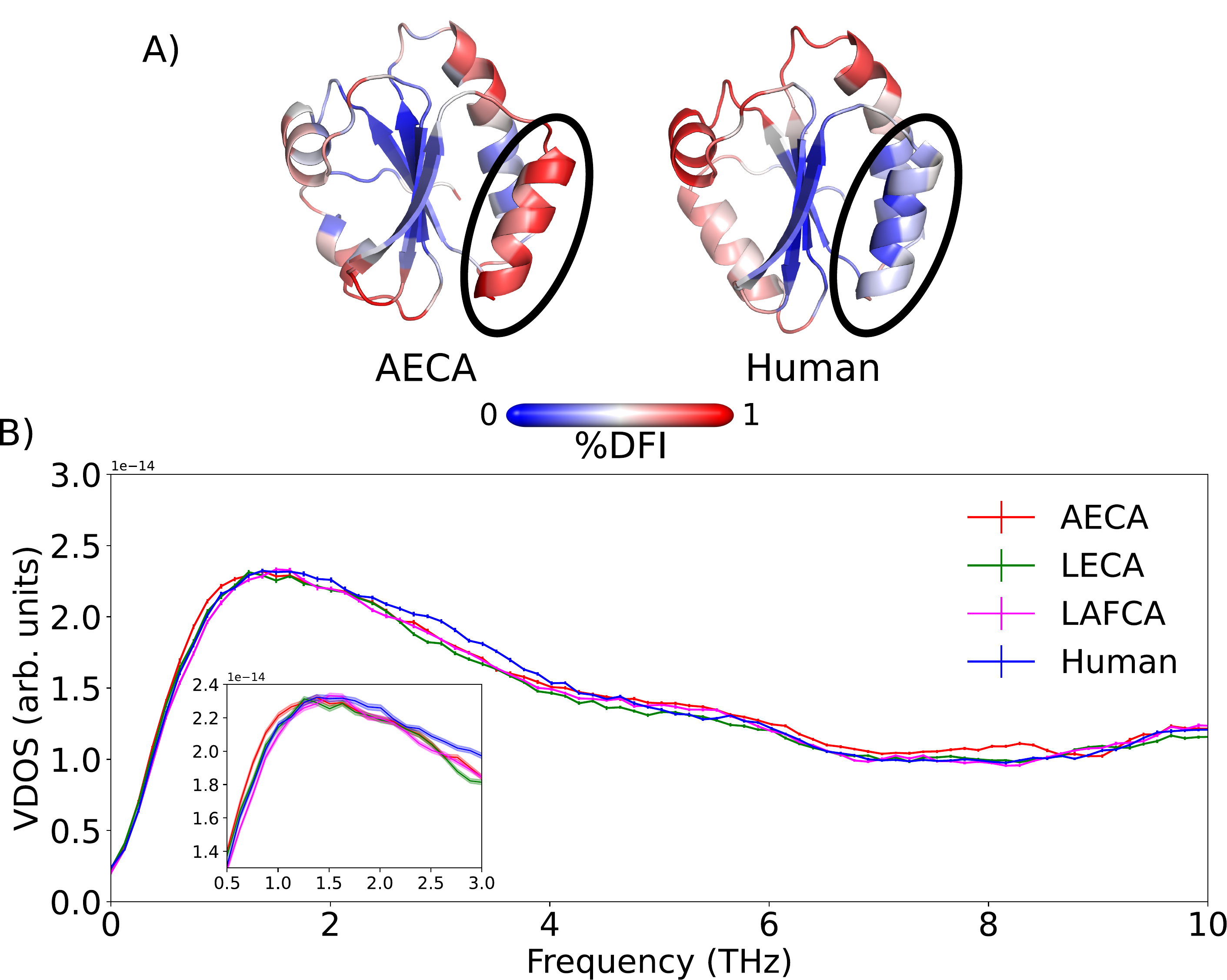}
    \renewcommand{\figurename}{Figure}
    \caption{\textbf{Modulation of the native state ensemble of Thrx enzymes from human branch observed using their VDOS.} (A) The change in  flexibility profile of the enzyme throughout the amino-acid substitutions as it  evolves to human Thrx enzyme. The ranked DFI profiles are mapped on 3-D structure as cartoon representations using a spectrum of red-white-blue. This change can also be observed by comparing the VDOS of the ancestral proteins with the extant ones (in (B)). The differences in lower frequencies are focused in the inset where the shaded region on the plots show the standard error of mean.}
\end{figure}

\begin{figure}[h!]
    \centering
    \includegraphics[width=8.6cm]{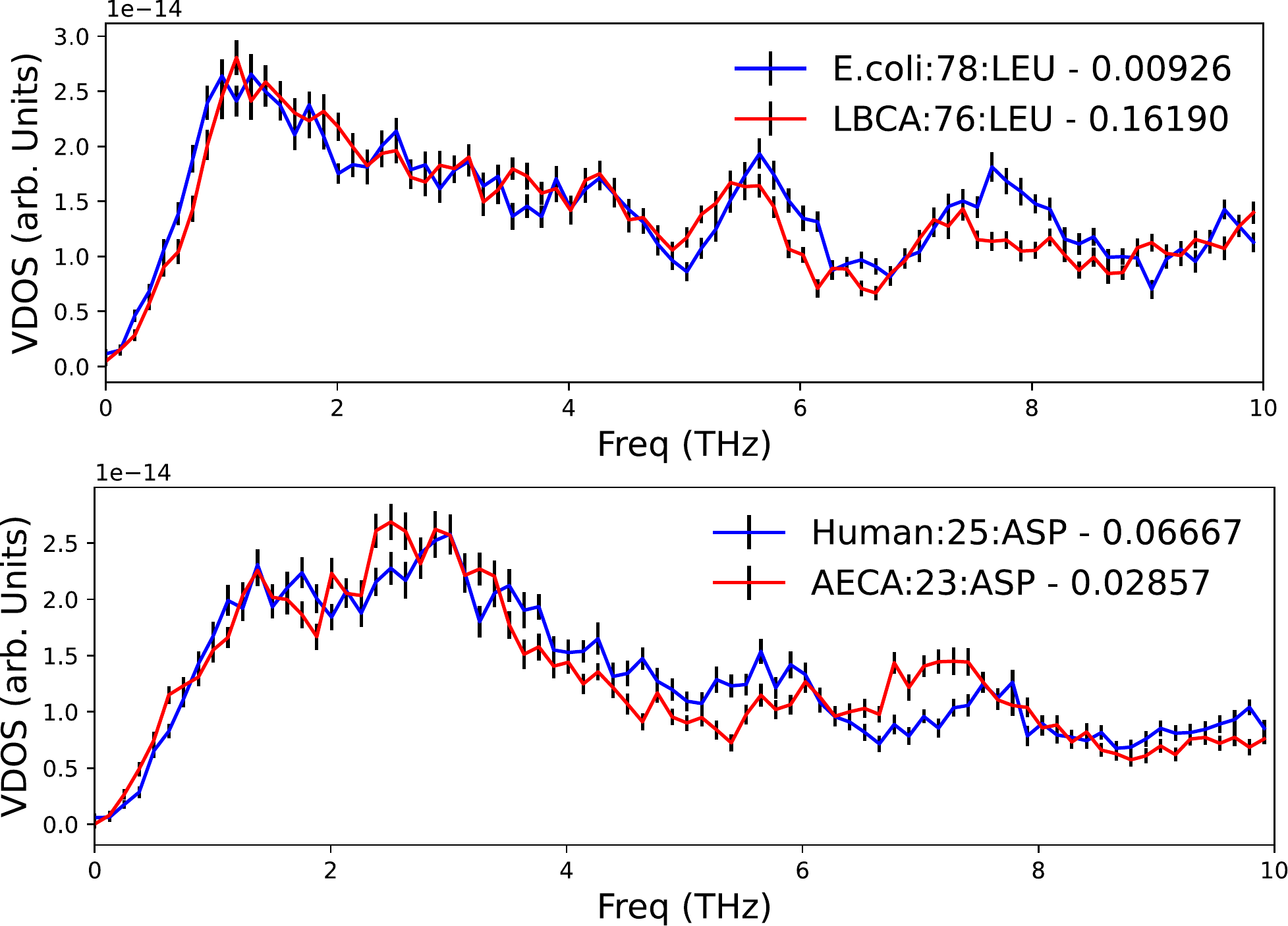}
    \caption{\textbf{Evolution preserves the VDOS of the residue positions with no substitution and no change in flexibility in Thrx proteins.} A subset of residues positions in Thrx proteins which have retained their rigidity as a hinge position (i.e., \%DFI $<$ 0.2) and also have not underwent any substitution show a higher degree of similarity between the VDOS of ancestral and extant enzymes from bacterial branch (top) and human branch (bottom). This indicates a similarity in slower and faster time scaled dynamics. The black ticks on plots indicate the standard error of mean.}
\end{figure}

\begin{figure}[h!]
\centering
\includegraphics[width=8.6cm]{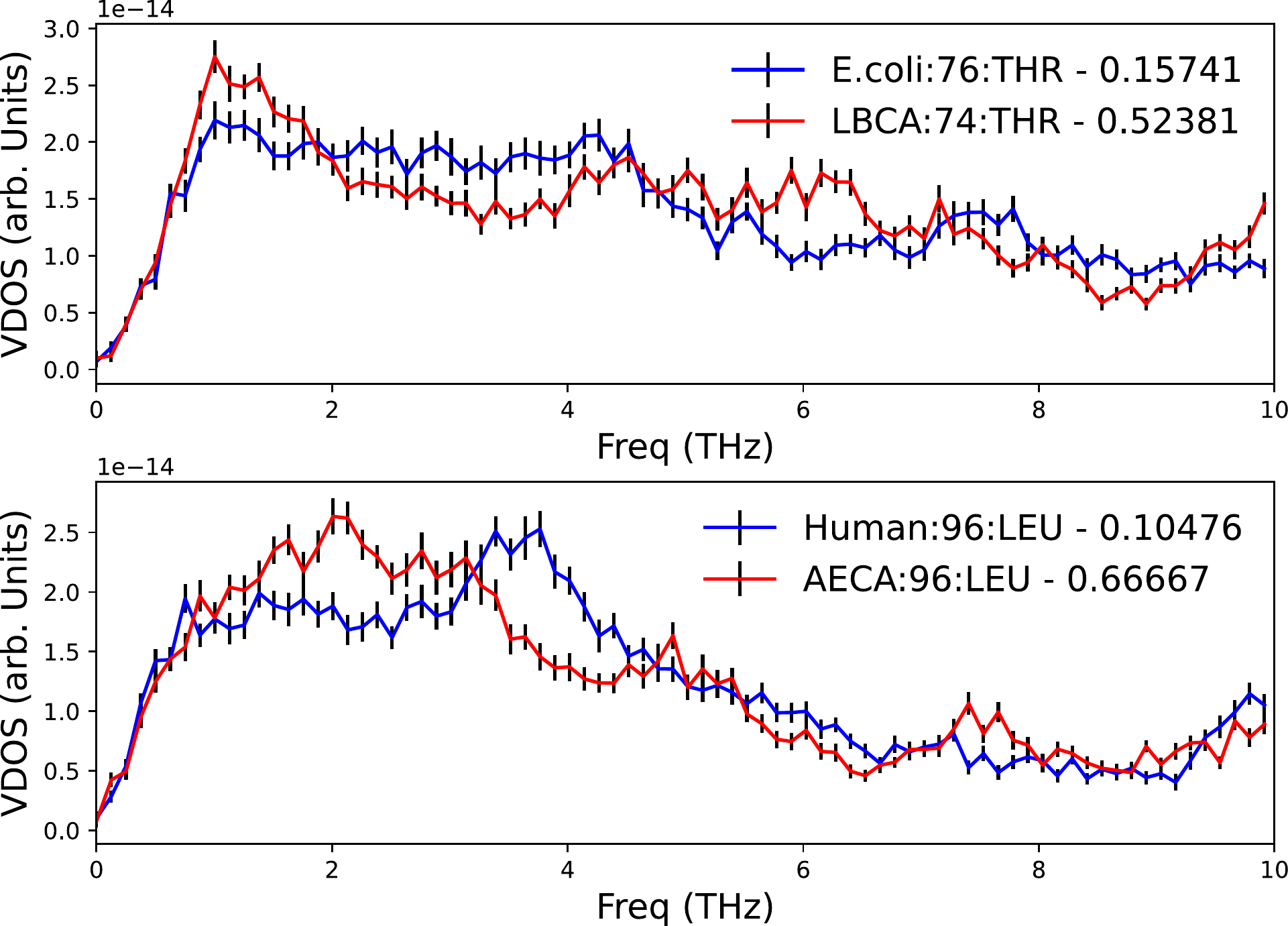}
\caption{\textbf{Change in VDOS of the residue positions where evolution with no substitution but a change flexibility through evolution in human and bacterial Thrx proteins.} Here we observed a subset of the residue positions in Thrx proteins with a conserved amino acid but a change in \%DFI score. Upon comparison of their VDOS revealed differences, particularly in lower frequencies where an increase in flexibility is accompanied by a shift of low frequency peaks towards even lower frequencies in bacterial Thrxs (top) and human Thrxs (bottom). The black ticks on plots indicate the standard error of mean.}
\end{figure}

\begin{figure}[h!]
    \centering
    \includegraphics[width=8.5cm]{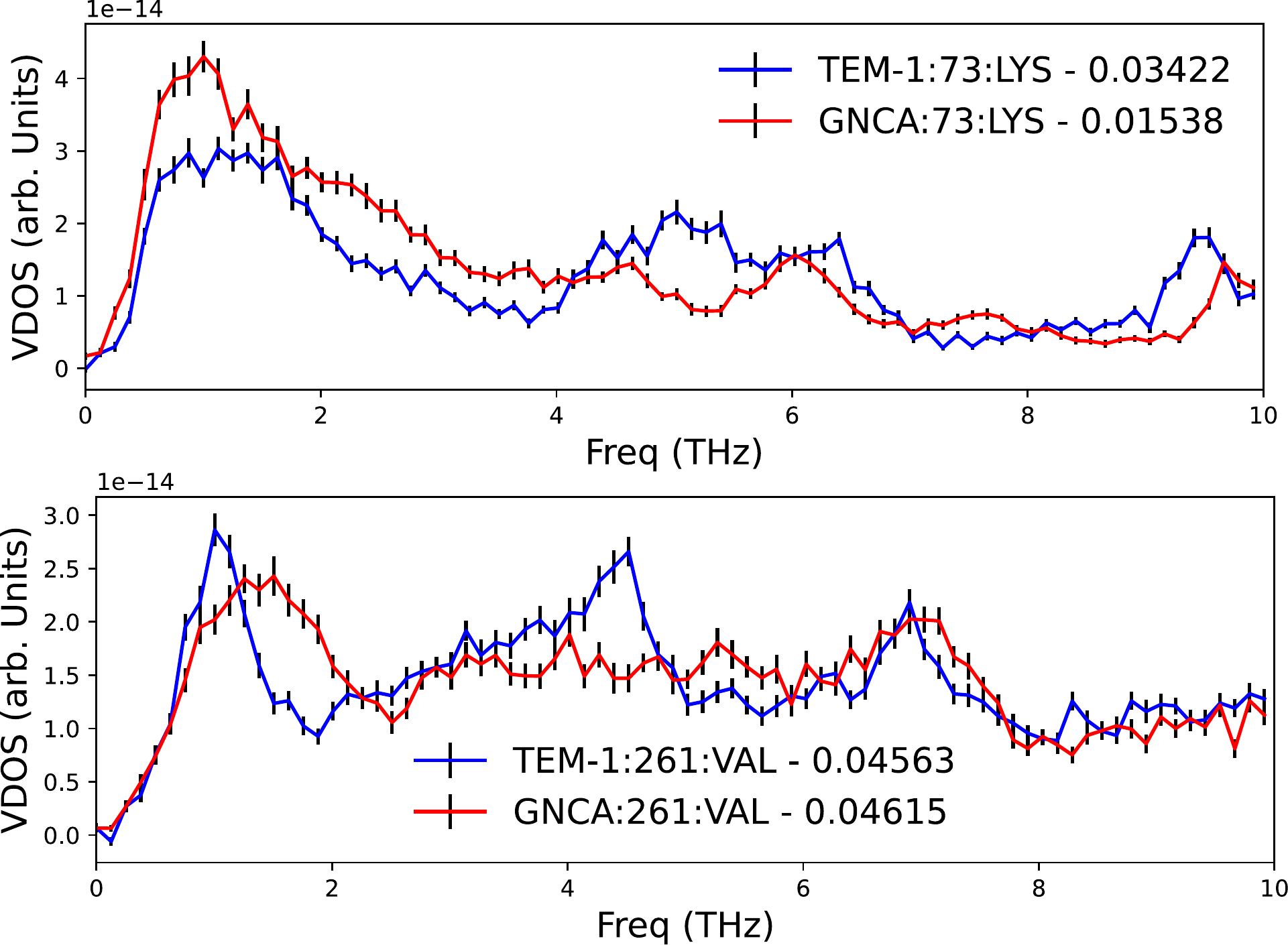}
    \caption{ \textbf{Lower frequencies in the VDOS are not always enough to describe the relevant dynamical changes.} Here we observed that, contrary to previous examples, taking into account only the lower frequencies do not provide enough information to represent changes in the dynamical flexibilities of all residue positions. This is shown here by a subset of examples of residue positions in $\beta$-lactamases which have retained their amino acid identity and rigidity through evolution but still exhibit significant differences in their VDOS, particularly at lower frequencies. The black ticks on plots indicate the standard error of mean.}
\end{figure}

\bibliography{manuscript.bib}
\bibliographystyle{ieeetr}